\documentclass[rsi, amsmath,amssymb,preprint]{revtex4-1}
\usepackage{comment}
\usepackage{graphicx}
\usepackage{dcolumn}
\usepackage{bm}

\usepackage[utf8]{inputenc}
\usepackage[T1]{fontenc}
\usepackage{mathptmx}
\usepackage{etoolbox}

\makeatletter
\def\@email#1#2{%
 \endgroup
 \patchcmd{\titleblock@produce}
  {\frontmatter@RRAPformat}
  {\frontmatter@RRAPformat{\produce@RRAP{*#1\href{mailto:#2}{#2}}}\frontmatter@RRAPformat}
  {}{}
}%
\makeatother
\begin{document}

\title[Single chamber MOT]{A compact setup for loading magneto-optical trap in ultrahigh vacuum environment }

\author{Kavish Bhardwaj$^{1}$}%
\author{Sourabh Sarkar$^{1, 2}$}
\author{S. P. Ram$^{1,*}$}
\author{V. B. Tiwari$^{1, 2}$}
\author{S. R. Mishra$^{1, 2}$}

\affiliation{$^{1}$Laser Physics Applications Section, Raja Ramanna Centre for Advanced Technology, Indore-452013, India.}
\affiliation{$^{2}$Homi Bhabha National Institute, Anushaktinagar, Mumbai-400094, India.}
    
\email{spram@rrcat.gov.in}


\begin{abstract}
We have developed a compact setup which enables loading a magneto-optical trap (MOT) in ultrahigh vacuum (UHV) environment. Nearly $1 \times 10^{8}$ atoms of $^{87}$Rb are trapped in the MOT at $\sim 2 \times 10^{-10}$ Torr base pressure in the chamber. After the MOT loading, we have successfully demonstrated working of quadrupole magnetic trap in this chamber with a lifetime of $\sim 8 $ s.  
\end{abstract}

\maketitle

\section{\label{sec:Intro}Introduction} A compact and reliable source of ultra-cold atoms is an essential requirement for several experimental investigations and device applications using cold atoms. To produce sample of ultra-cold atoms, the cold atoms trapped in a magneto-optical trap (MOT) \cite{MOTPRL1987,BookInguscio2013,MOTcooling1991} are subjected to evaporative cooling \cite{Mishra2017} in an ultrahigh vacuum (UHV) environment. But, loading a MOT in ultrahigh vacuum (UHV) chamber has always been a complicated task to deal with, which requires either a double MOT setup \cite{S_P_RAM_Double_Chamber_MOT_2013} or a zeeman slower device \cite{Slowe_Zeeman_Slower_2005}. In first method (i.e. use of double-MOT setup), a UHV chamber MOT (UHV-MOT) is loaded by pushing the cold atom from a vapor chamber MOT to UHV chamber through a narrow  differential pumping tube connecting two chambers. In Zeeman slower method to load MOT in UHV environment, a thermal atomic beam is injected in UHV chamber which is slowed down by using a Zeeman slower device for UHV-MOT loading. Both these methods are complicated in one or the other way. Hence, a simpler method to load MOT in UHV environment is necessary and useful. 

In this paper, we present a simple design to prepare and load MOT in UHV chamber from thermal source of atoms (from Rb-dispenser). We demonstrated that, even after the MOT loading, the vacuum of the chamber is preserved to the extent that magnetic trapping of atoms in the same chamber can be performed with good life-time of the trap. The observed life-time of the quadrupole nagnetic trap was $\sim 8 s $, which may be further improved by optimizing the current pulse parameters applied to Rb-dispensers. Such simple and portable cold atom setups may be useful for developing several atom-optic devices for magnetometry \cite{OpticalMagnetometryNature2007}, atom interferometry \cite{AtomInterferometry2009} and other quantum technologies\cite{QuantumTechnologies2003}. 

\section{\label{sec:expt}Experimental}
\begin{figure}
\includegraphics[width=0.7\textwidth]{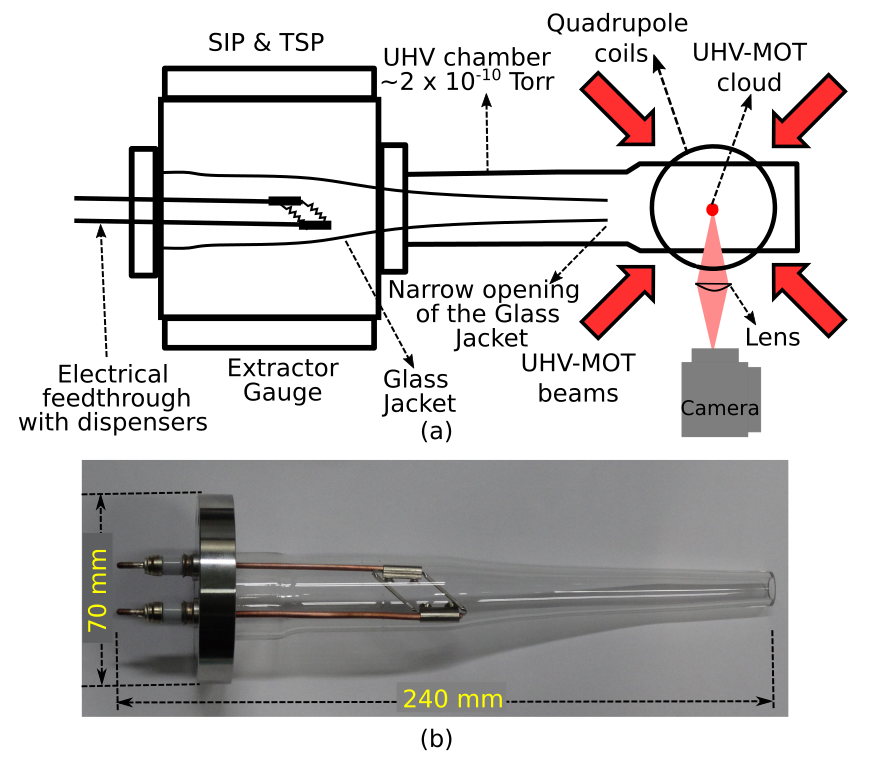}
\caption{\label{fig:Schematic} (a): Schematic of the setup with glass jacket for formation of magneto-optical trap (MOT) in ultrahigh vacuum environment. Two pairs of MOT laser beams (out of three) are shown in the figure and the third pair of MOT beams is perpendicular to plane of paper. The MOT is formed in the UHV environment by supplying a current to the feedthrough and (b): dispenser assembly inside the glass jacket.}
\end{figure}

The setup is shown schematically in Fig.\ref{fig:Schematic}. 
We have used a stainless steel (SS) cube as UHV chamber. At one end of this SS cube, a UHV compatible square cross-section glass cell is attached (Fig. 1) using a CF35 flange. The glass cell has clear optical access region of nearly $70$ mm $x$ 30 mm $x$ 30 mm. The UHV-MOT and magnetic trap can be formed in this glass cell. The glass cell is surrounded by the coils which are used to generate the required magnetic field for MOT and magnetic trap. At other end of the SS cube, two Rb-dispensers (Rb/NF/4.8/17FT SAES getters) are inserted via a two-pin UHV feethrough of length 70 mm inside the chamber. These Rb-dispensers are surrounded by a hollow glass jacket made of 2 mm thick Borosil glass having a length of 200 mm. The jacket has a tapered opening of diameter 8 mm towards the UHV glass cell. The base diameter of the jacket is 28 mm. The jacket is push-fit attached to feedthrough flange (CF35) with the use of a spring coil at glass cell flange. The hot atoms emitted by dispensers strike the glass jacket walls multiple times before reaching the  glass cell kept at UHV pressure of $\sim 2 \times 10^{-10}$ Torr. Thus atoms from dispenser are slowed down by the walls of glass jacket before being trapped in MOT formed in the glass cell. Earlier studies have also reported loading of MOT in UHV environment by using more sophisticated cooling devices for hot atoms emitted from dispensers \cite{AtomicbeamMOT2004,PulsedMOT2005}. The glass jacket which we use is simpler and a passive device which can slow down hot atoms and  redirect them towards UHV-MOT region. 

A pressure of $\sim 2 \times 10^{-10}$ Torr is generated in the glass cell using a turbo-molecular pump (TMP), a sputter ion pump (SIP) and a titanium sublimation pump (TSP). The whole setup was baked at $150 ^{o}C$ for 24 hours and de-gasing of dispensers and activation of TSP has been carried out for achieving this UHV environment. The Rb vapor is generated in the UHV chamber by applying a DC current through the dispensers. Here, two Rb-dispensers are kept in parallel arrangement. This parallel arrangement provides direct contact of each dispenser to copper feedthrough rods to have a better conductance for heat dissipation. In our setup, expected dispenser temperature is $\sim 500 ^{o}$ C (estimated from the data sheet of the dispenser at 4 A current) during the MOT loading. After MOT loading, dispenser temperature is to be brought down quickly so that MOT and magnetic trap loading is minimally disturbed by hot atoms from dispensers. Neglecting radiative cooling, the time required for the dispenser to cool down is inversely proportional to conductance. Atoms are emitted from the dispenser by heating the dispenser by flowing $8$ A current in dispenser assembly. The atoms are finally collimated towards glass cell via narrow opening in the glass jacket for trapping in MOT \cite{AtomicbeamMOT2012}.

For formation of MOT, two ECDL laser (TOPTICA DL-100) operating at wavelength $780$ nm and a laser amplifier (TOPTICA, Boosta) provides the necessary laser requirement. Three orthogonal pairs of counter-propagating laser beams having intensity 22 mW/cm$^2$ and frequency of 15 MHz red-detuned to cooling transition ($^{5}S_{1/2}$ F = 2 to $^{5}P_{3/2}$ F$^\prime$ = 3) of $^{87}$Rb are made intersecting one another inside the glass cell. A quadrupole magnetic field having axial field gradient of 10 G/cm is used for MOT formation. Along with the cooling laser beams, we apply another pair of counter propagating re-pumping laser beam having frequency resonant to re-pumping transition ( $^{5}S_{1/2}$ F = 1 to $^{5}P_{3/2}$ F$^\prime$ = 2) of $^{87}$Rb. The MOT loading was studied by collecting fluorescence from the atoms in MOT on a calibrated photodiode.

After loading the MOT in this setup, we have also performed the magnetic trapping of atoms. For loading magnetic trap, the laser beams were switched-off and quadrupole magnetic field gradient was raised from ~10 G/cm to 220 G/cm. For this, the current was flown through quadrupole magnetic coils through IGBT based current switch and was ramped to 22 A in 500 ms using a c-RIO based electronic controller. The number of atoms in magnetic trap was estimating by fluorescence imaging of the atom cloud in trap. 

\section{\label{sec:Results}Results and discussion}

For the study the loading behavior of MOT in the UHV environment, we have collected the fluorescence from MOT atom cloud on a calibrated photodiode. Figure 2 shows the photodiode signal due to fluorescence from the atoms in the MOT. The photodiode signal is proportional to number of atoms in the MOT. Final equilibrium number of atoms trapped in the MOT (N) is determined by the balance between capture rate (R) and loss rate of the MOT. The MOT loading was performed in continuous and pulsed mode operations of current in dispenser source. 

\begin{figure}
\includegraphics[width=\textwidth]{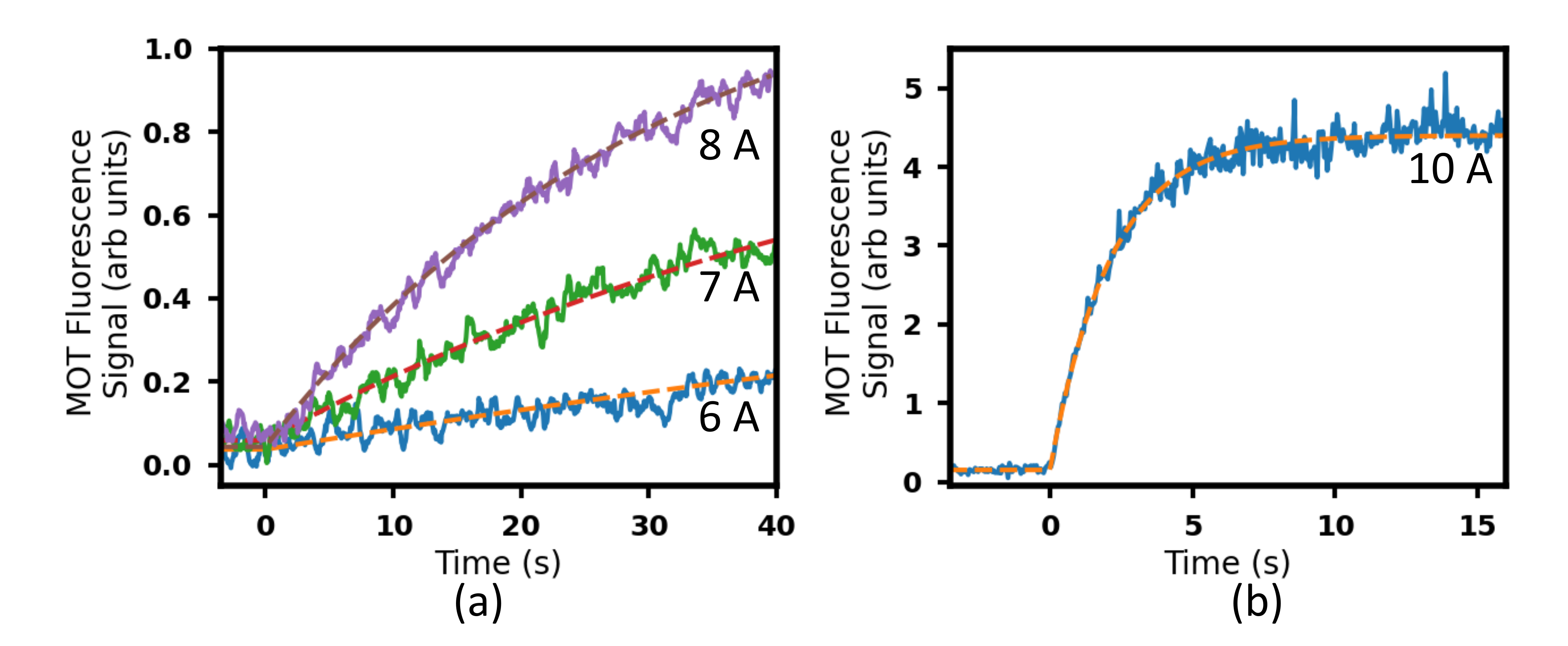}
\caption{\label{fig:Loadingcurves} (Color online) Observed loading  curves of UHV-MOT for different values of dispenser currents in continuous mode, (a) for 6 A, 7 A, and 8 A current in dispenser assembly and (b) for 10 A current in dispenser assembly. In plots the experimental data are shown by continuous curves whereas the fitted data are shown by dashed curves. }
\end{figure}

\begin{figure}
\includegraphics[width=0.8\textwidth]{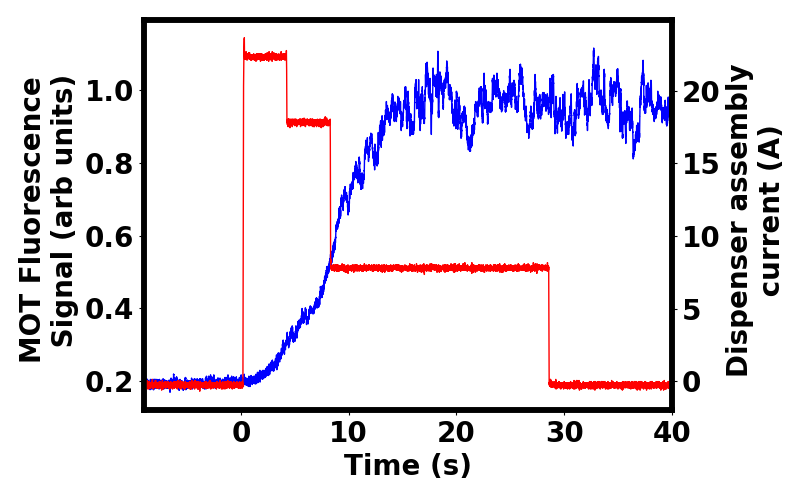}
\caption{\label{fig:PulsedLoading}(Color online) MOT loading curve (blue) with a structured current pulse (red) in the dispenser assembly.}
\end{figure}

Figure \ref{fig:Loadingcurves} shows the loading curves of $^{87}$Rb UHV-MOT for different values of current in dispensers assembly. Figure \ref{fig:Loadingcurves} (a) is for 6 A, 7 A, and 8 A values of current in dispenser assembly. The loading time $\tau$ estimated from fitting of the experimental data to the equation $N = N_{s}(1 - e^{(-t/\tau)})$ are 150 s, 59 s and 31 s respectively for the loading curves corresponding to 6 A, 7 A and 8 A current in dispenser assembly. Here, $N$ is the instantaneous number of atoms in the MOT,  $N_{s}$ is the saturated number in the MOT and $t$ is the time. At 8 A current in the dispenser assembly, $\sim$ $7.5 \times 10^7$ atoms are obtained in the UHV-MOT. At this value of dispenser current (8 A in continuous mode),  the vacuum inside chamber deteriorates and the pressure reaches to a steady state value of $\sim 5.5 \times 10^{-10}$ Torr from the initial value of $\sim 2 \times 10^{-10}$ Torr. At further higher current, e.g. 10 A in continuous mode, the loading time reduces to few seconds (Fig. \ref{fig:Loadingcurves}(b)) and number of atoms increases to $\sim$ $2.5 \times 10^8$, but the vacuum in the chamber deteriorates very quickly (pressure rising to $\sim 8 \times 10^{-10}$ Torr in few tens of second). 
In order to have faster loading and less deterioration of vacuum in the chamber, we tried to load the MOT using a structured current pulse in the dispenser assembly.
A structured current pulse used in the experiments is shown in Fig. \ref{fig:PulsedLoading} by red colour curve. In the pulse structure, the initial part of the pulse has high current for short duration and later part of the pulse has low current for long duration. With this structure of current pulse, the MOT loading time reduces to $\sim 9 s$ and vacuum is recovered at the end of the current pulse. The MOT loading is shown by the blue curve in Fig. \ref{fig:PulsedLoading} using this current pulse. The pulsed loading sequence used here, leads to the loading of $1 \times 10^8$ atoms in the MOT.

A major difference in MOT loading with structured current pulse and continuous current mode comes from the use of high current in structured pulse which increases the temperature of the dispensers. Therefore, if the next MOT loading cycle is repeated soon without allowing the cooling of dispensers, the temperature of the dispensers increases in accumulative way with successive MOT loading cycles. This leads to accumulative increase of pressure in the chamber due to increased Rb emission. Similar observations have been reported earlier by  \citet{P_Griffin_Fast_Switching_2005}. Therefore, this type of MOT loading can be preferred with low repetition rate of MOT loading.    

\begin{figure}
\includegraphics[width=11 cm]{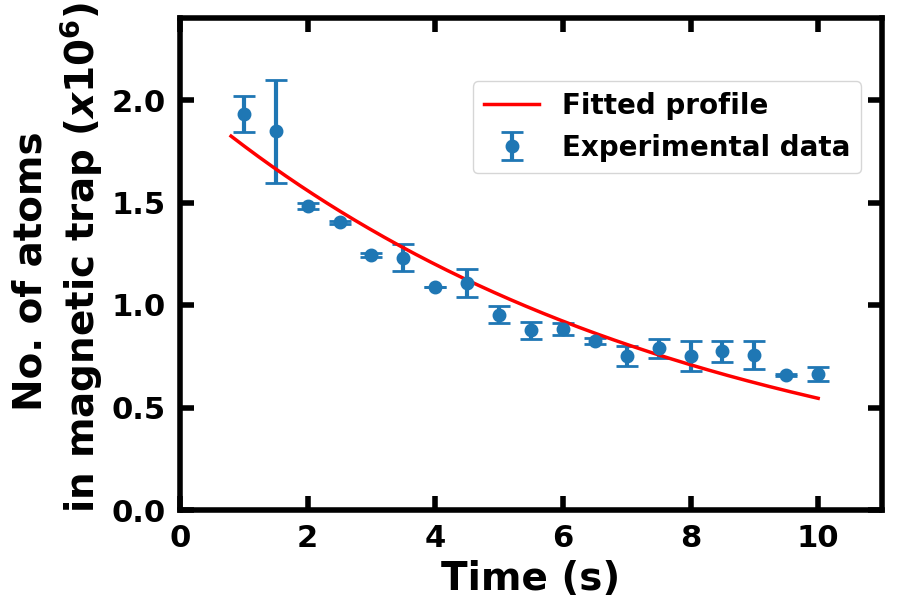}
\caption{\label{fig:Lifetime} Measured variation (circles) in number of atoms in quadrupole magnetic trap with time. The continuous curve is an exponential decay curve fitted to estimate the lifetime of the atoms in the magnetic trap.}
\end{figure}

After successful loading of MOT in UHV environment, we demonstrated the use of the UHV-MOT for magnetic trapping of atoms. We transferred the cold $^{87}$Rb atoms from the MOT to a quadrupole magnetic trap formed by two coils. From the variation in number of atoms in the magnetic trap with time (as shown in Fig. \ref{fig:Lifetime}), the $1/e$ lifetime of the trap was $\sim$ 8 s when this MOT was loaded with 8 A current (continuous) in the dispenser assembly. The dispenser current was continuously ON during the magnetic trapping of the atoms. A good lifetime of the magnetic trap indicates that a high vacuum in the chamber is maintained. This lifetime can be further improved by taking various measures in magnetic trapping such as blocking the residual lights going to the chamber, blocking the Majorana losses, etc, \cite{Mishra2017}.  A good lifetime of the atoms in the magnetic trap can be exploited for further experiments such as evaporative cooling \cite{Mishra2017}, radio-frequency (RF) dressing \cite{Chakraborty_2016}, Bose-Einstein condensation \cite{Mishra2017, Van_der_Stam_Na_BEC_2007}, etc, in the same chamber. 
\section{\label{sec:Conclusion}Conclusion}
We have developed a compact setup and used it for magneto-optical trap (MOT) loading in ultrahigh vacuum (UHV) environment. The magnetic trapping of atoms using this UHV-MOT is also demonstrated in this compact setup. 

\begin{acknowledgments}
We  thank S. Singh and V. Singh for useful technical discussions. We are thankful to Shri Deepak Sharma and Shri Bhupinder Singh for the development of switching circuit, Shri Ayukt Pathak and Smt. Shraddha Tiwari for the development of c-RIO based electronic controller and Shri Ajay Kak for manufacturing glass jacket. 
\end{acknowledgments}

\section*{Data Availability Statement}
Data available on request from the authors.

\providecommand{\noopsort}[1]{}\providecommand{\singleletter}[1]{#1}%

\end{document}